\documentclass[aps,prd,preprint,groupedaddress,nofootinbib]{revtex4}
\usepackage[mathscr]{euscript}
\usepackage{mathtools}
\usepackage{graphicx}
\usepackage{float,epsfig}
\usepackage{bm}
\usepackage{graphicx}
\usepackage{subfigure}
\usepackage[colorlinks=true,linkcolor=blue,urlcolor=blue]{hyperref}
\newcommand{\bea}{\begin{eqnarray}}
\newcommand{\eea}{\end{eqnarray}}
\newcommand{\beq}{\begin{equation}}
\newcommand{\eeq}{\end{equation}}

\def\/{\over}

\begin{document}

\title{Thermodynamical multihair and phase transitions of
4-dimensional charged Taub-NUT-AdS spacetimes}

\author{Yao Xiao$^{1}$, Jialin Zhang$^{1}$\footnote{Corresponding author at jialinzhang@hunnu.edu.cn}, Hongwei Yu$^{1}$\footnote{Corresponding author at hwyu@hunnu.edu.cn} }

\affiliation{$^{1}$Department of Physics and Synergetic Innovation Center for Quantum Effects and Applications, Hunan Normal University, Changsha, Hunan 410081, China}

\begin{abstract}
We study the behavior of phase transitions for the four-dimensional
charged Taub-NUT-AdS spacetime  with the Newman-Unti-Tamburino (NUT)
parameter interpreted  as the thermodynamic multihair in the extended
thermodynamic phase space, and mainly focus on the effects of the NUT
parameter  on the  phase transitions. We find that there is an upper bound on the value of the NUT parameter  beyond which the corresponding physical inflection point or critical point will not exist, and the  thermodynamic trihair interpretation of the
NUT parameter would admit a little larger upper bound than the thermodynamic bihair
interpretation. Moreover, as long as  the NUT parameter is
vanishingly small, the analogy to the van der Waals liquid/gas phase
transition is valid irrespective of the  multihair characteristics
of the NUT parameter. However, as the NUT parameter increases to be
comparable to the electric charge,  such analogy to the van der Waals
system will be broken, and the corresponding inflection point is not a  thermodynamic critical point any more. For a large NUT parameter,  there are frequent
occurrences of the zeroth order phase transition  in the case of the
thermodynamic bihair interpretation,  while only the first
order phase transition happens in the case of the thermodynamic trihair
interpretation.
\end{abstract}

\maketitle

\section{Introduction}
\label{sec_in} \setcounter{equation}{0} It has been more than  four
decades since the concept of  black hole entropy was well established in
the pioneering work of  Bekenstein and
Hawking~\cite{Bekenstein:1973,Hawking:1976}. Since then, a wide
variety of phenomena concerning the thermodynamic aspects of black
holes have been studied, which  provide  deep  insights into the
quantum nature of gravity. It was found that black
holes can exhibit some incredibly rich thermal phase structures and
phase transitions between different spacetime geometries.
Particularly, for black holes in  Anti-de Sitter (AdS) space,  which
are thermodynamically stable due to fact that AdS space acts as
a confined cavity, their thermodynamical properties are quite
different from those  in asymptotically flat or de Sitter space.
The first study about the thermodynamics of AdS black holes was
reported by Hawking and Page~\cite{Page:1983}, who  found  that
there could be a phase transition between the stable large black
hole and the thermal gas  for Schwarzschild black holes in  AdS
space. It is worth pointing out that the thermodynamic behavior of black
holes in AdS space is often utilized  in strongly coupled  gauge
theories by means of AdS/CFT
correspondence~\cite{Maldacena:1998,Gubser:1998,Witten:1998-1,Witten:1998-2,Aharony:2000}

More recently, motivated by the basic thermodynamic scaling
argument, it was realized that the negative cosmological constant
should be interpreted  as the pressure in the extended black hole
thermodynamics and its conjugate quantity is just  the thermodynamic
volume~\cite{Mann:2012}.  With this proposal,  some
interesting thermodynamic phenomena and rich phase structures quite
analogous to that of the van der Waals fluid are discovered for charged AdS
black holes~\cite{Mann:2012,Chamblin:1999,Gunasekaran:2012}, e.g.,
the liquid/gas transition for the van der Waals fluid is analogous
to the small/large black hole phase transition. A lot of more interesting phase transition
phenomena for black holes in AdS space have  been demonstrated as well,
including triple points and reentrant phase
transitions~\cite{Altamirano:2013,Altamirano:2014}, multiple
re-entrant phase transitions~\cite{Frassino:2014}, isolated critical
points~\cite{Dolan:2014,Hennigar:2015} and a super fluid phase
transition~\cite{Hennigar:2017}. These thermodynamic phenomena exhibit  novel chemical-type
phase behaviors, leading to the naming of black hole chemistry
~\cite{Mann:2014,Mann:2017}.

On the other hand, asymptotically locally flat spacetimes endowed with a nonzero NUT charge posed a great challenge to the study of thermodynamics.
The Taub-Newman-Unti-Tamburino (Taub-NUT) solution as a simple non-radiating exact solution to
Einstein's field equations was firstly discovered by Taub in
1951~\cite{Taub:1951}, and subsequently rediscovered  as a candidate
for a black hole in the generalization of the Schwarzschild
space-time~\cite{Newman:1963}. The Lorentzian Taub-NUT metric is one
of the most intriguing solutions since it features a string like
singularity (usually called Misner string singularity ) on the polar axis with closed
timelike curves in its vicinity, and carries  a peculiar type
of the NUT charge (the Misner gravitational charge) which is  somewhat
analogous to the magnetic monopole.  In order to avoid these
singularities, Misner once imposed a periodical condition for the
time coordinate in the Lorentzian Taub-NUT metric
~\cite{Misner:1963}.  Such a periodical identification,  however,
may create closed timelike curves everywhere, rendering  the maximal
extension of the spacetime problematic~\cite{Hajicek:1971,Hawking:1973}.

Recently, it has been argued that the Taub-NUT-type spacetime with
the presence of Misner strings is less pathological than
previously thought.  The Misner strings are
transparent for geodesics, which could render the spacetime
geodesically  complete  and be free of causal pathologies for
freely falling
observers~\cite{Miller:1971,Clement:2015,Clement:2016,Mann:2019}.
Based on these arguments, a lot of efforts have been made to
formulate a consistent and reasonable thermodynamical first law for
the Lorentzian Taub-NUT-type spacetimes with the presence of Misner
strings~\cite{Mann:2019,Bordo:2019,Bordo:2019-2,Ballon:2019,Durka2019}.
In such developments, a pair of new conjugate quantities related
to the NUT parameter are introduced as a crucial step to obtain the
first law. However, it has been argued in Ref.~\cite{Wu:2019} that
the thermodynamical  quantity associated the  NUT parameter  in the
aforementioned  studies of the thermodynamics  does not possess the
conventional characters of global charges measured at infinity.

Therefore, it was suggested that the NUT parameter should
be considered as a thermodynamic multihair rather than a single feature of
the physical source, and then the appropriate thermodynamic first
law for Taub-NUT-type spacetimes could be naturally formulated by
first deriving the Christodoulou-Ruffini-type squared-mass formula without
imposing any constraint condition.

 In this paper, based on  the thermodynamical first law
derived in Ref.~\cite{Wu:2019}, we perform a study on the behavior
of possible phase transitions for the 4-dimensional Lorentzian
Reissner-Nordstr$\rm\ddot{o}$m-Taub-NUT-AdS (RN-Taub-NUT-AdS)
spacetime in the extended thermodynamical phase space, and examine the
analogy to the van der Waals-like liquid/gas transition.  We also
attempt to seek the influence of the NUT parameter as a poly-facet on
the corresponding phase structure and the phase transition and
make a cross-comparison of the  role of thermodynamic bihair and
trihair played by the NUT parameter in phase transitions. The
organization of the paper is as follows. In next section, we will
review the  thermodynamics for the  RN-Taub-NUT-AdS
spacetime with the NUT parameter interpreted as  a thermodynamic
bihair (angular momentum and NUT charge) or a thermodynamic trihair
(angular momentum, NUT charge and magnetic mass).  In Sec.III, we
investigate the influence of the NUT parameter on the phase
transitions with the NUT parameter interpreted as a thermodynamic
bihair and  thermodynamic trihair, respectively. This allows  a
comparison between the behaviors of phase transitions for the $N$-bihair
solution and the $N$-trihair solution.  Meanwhile, the presence  of
the critical point or inflection point is also discussed in detail.
 Finally, we end up in Sec.IV by summarizing the results.

\section{The first law with the NUT parameter interpreted as a thermodynamic multihair} \label{sec_ba}
For the 4-dimensional RN-Taub-NUT-AdS spacetime
with a nonzero cosmological constant and a pure electric charge, the
metric can be written in following form with the Misner strings
symmetrically distributed along the polar
axis~\cite{Newman:1963,Wu:2019,Alonso:2000,Mann:2006,Awad:2006,Johnson:2014}
\begin{equation}\label{metric1}
ds^2=-\frac{f(r)}{r^2+N^2}(dt+2N\cos\theta
d\phi)^2+\frac{r^2+N^2}{f(r)}dr^2+(r^2+N^2)(d\theta^2+\sin^2\theta
d\phi^2)\;,
\end{equation}
where
\begin{equation}
f(r)={r^2-2Mr-N^2+Q^2}+\frac{(r^4+6N^2r^2-3N^4)}{\ell^2}
\end{equation}
and the one-form electromagnetic  potential
\begin{equation}
A=\frac{Q r}{r^2+N^2}(dt+2N \cos\theta d\phi)\;.
\end{equation}
Here,  the parameters $M$, $Q$, $N$ and $\ell$ represent  the
electric mass, the electric charge, the NUT charge and the AdS
radius, respectively. In the discussions which follow, we  only
consider the case in which the charges, i.e.,  $Q$ and $N$, are all
positive for simplicity. Introducing $J_N=NM$ as the ``angular
momentum", the second physical feature of the NUT parameter,  and
after some algebraic   manipulations, the Christodoulou-Ruffini-type
squared-mass formula~\cite{Christodoulou:1970} can be derived
straightforwardly
\begin{equation}\label{mass-sqr}
M^2=\frac{\pi}{4S}\bigg[\Big(1+\frac{32\pi}{3}P
N^2\Big)\Big(\frac{S}{\pi}-2N^2\Big)+Q^2+\frac{8S^2P}{3\pi}\bigg]^2+\frac{J_N^2\pi}{S}\;,
\end{equation}
where the thermodynamic  pressure $P={3}/{(8\pi \ell^2)}$ and the entropy $S=\pi(r_+^2+N^2)$ with $r_+$ representing  the radius of the outer  horizon.

The differentiation of Eq.~(\ref{mass-sqr}) leads to  the first
thermodynamic law with the NUT parameter interpreted as a
thermodynamic bihair ($N$-bihair solution)~\cite{Wu:2019}
\begin{equation}
dM=TdS+\omega_h dJ_N+\Psi_h dN+\Phi{dQ}+VdP,
\end{equation}
where the corresponding conjugate thermodynamic variables satisfy
\begin{align}\label{var-all1}
T&=\frac{f'(r_+)}{4\pi(r_+^2+N^2)}=\frac{r_+-M+2r_+(r_+^2+3N^2)/\ell^2}{2\pi(r_+^2+N^2)}\;,\qquad\Phi=\frac{Qr_+}{r_+^2+N^2}\;,\nonumber\\
\Psi_h&= \frac{4Nr_+(r_+^2-3N^2)-2Nr_+\ell^2}{(r_+^2+N^2)\ell}\;,\qquad\qquad\qquad\omega_h=\frac{N}{r_+^2+N^2}\;,\nonumber\\
V&=\frac{4\pi r_+(r_+^4+6N^2r_+^2-3N^4)}{3(r_+^2+N^2)}\;.
\end{align}
Here,  the value
of the NUT parameter $N$ can not be overwhelmingly large in comparison
with $Q$ or $r_+$ so as to hold the  thermodynamic volume positive.  In other words, the horizon radius $r_+$ should be larger than $\sqrt{2\sqrt{3}-3}N$  to meet  the requirement of a positive volume. In the following discussions, we will assume that this condition is satisfied, in accordance with the usual analysis of  the van der Waals liquid-gas system.
If a positive thermodynamic volume is assumed,
one can choose to interpret
the NUT parameter $N$  as the thermodynamic trihair, i.e, the angular
momentum, NUT charge and magnetic mass. Thus, in the metric
described by Eq.~(\ref{metric1}), another consistent first law of
thermodynamics can also be obtained by assuming angular momentum
$J_N=N M$ and magnetic mass $\widetilde{M}=N(1+4N^2/\ell^2)$. Note here that the positivity of $N$ ensures the positivity of $\widetilde{M}$.
Through some algebraical manipulations,  the thermodynamic volume
can be derived as $\widetilde{V}={4\pi r_+(r_+^2+3N^2)}/{3}$ via
appropriately adjusting the identity $f(r_+)=0$ and the expression
of the square-mass formula. Concretely, the corresponding first law  for the
$N$-trihair solution can be written as~\cite{Wu:2019}
 \begin{equation}
dM=TdS+\omega_h dJ_N +\Phi{dQ}+\widetilde{\Psi}_h
dN+\xi{d\widetilde{M}}+\widetilde{V}dP,
\end{equation}
where the additional or new conjugate thermodynamic quantities
satisfy
\begin{align}\label{var-all2}
\widetilde{\Psi}_h=-\frac{2Nr_+}{r_+^2+N^2}+\frac{4N^2-\ell^2}{\ell^2}\xi\;,\qquad\xi=\frac{r_+(r_+^2-3N^2)}{4N(r_+^2+N^2)}\;.
\end{align}
 It is worth pointing out here that the  expression of the
thermodynamic volume $\widetilde{V}={4\pi r_+(r_+^2+3N^2)}/{3}$ is
completely identical to
 the result given in Refs.~\cite{Bordo:2019,Bordo:2019-2,Mann:2019} where the first law  is formulated by introducing a new
thermodynamic  NUT charge together with its conjugate quantity that
is not  a global conventional charge but a combination of parameters
$N$, $Q$, $\ell$ and $r_+$ . It is easy to find that no matter how large the NUT parameter will
be, the thermodynamic volume $\widetilde{V}$  is always nonnegative.
In comparison with the case of the $N$-bihair solution, the thermodynamic volume of the $N$-trihair solution approaches  zero only in the condition $r_+=0$.

To gain a better understanding of the influence of the NUT parameter on the thermodynamic properties of the RN-Taub-NUT-AdS spacetime, we will next discuss the possible phase structure and the phase transition in the cases of  the $N$-bihair solution and $N$-trihair solution respectively.
\section{The phase transitions with the NUT parameter being a thermodynamic multihair} \label{sec_ba2}

In  above discussions, the equation of state  and the
corresponding  thermodynamic quantities have  been derived
explicitly.  We  are now going to  explore the influence of
the NUT parameter on the phase structure of  the 4-dimensional charged
Taub-NUT-AdS spacetimes, and examine the analogy to the van der Waals
liquid/gas transition.

\subsection{$N$-bihair solution}

It is easy to find that when the NUT parameter vanishes the
metric~(\ref{metric1}) reduces to the 4-dimensional charged AdS black
holes. For a nonzero NUT parameter, Eq.~(\ref{var-all1}) can be translated into the
equation of state for the charged Taub-NUT-AdS spacetime  with fixed $Q$
and $N$, that is
\begin{equation}\label{state-1}
P=\frac{r_+T}{2(N^2+r_+^2)}-\frac{1}{8\pi(N^2+r_+^2)}+\frac{Q^2}{8\pi(N^2+r_+^2)^2}\;,
V=\frac{4\pi r_+(r_+^4+6N^2r_+^2-3N^4)}{3(r_+^2+N^2)}\;.
\end{equation}

Recalling that  the corresponding critical point is determined by  the
inflection point in the pressure-volume diagram, we have
\begin{equation} \label{cri-con}
\frac{\partial P}{\partial V}=0,\qquad \frac{\partial^2 P}{\partial
V^2}=0.
\end{equation}
To  facilitate  the  study on the possible behavior of  phase transitions in the pressure-volume diagram,  the nonnegative $V^{1/3}$  can be approximately
treated as  the specific volume in the following discussions due to
$V^{1/3}\propto(r_+^2+5N^2/3)/r_+$ for a not too large value of $N$.

 Since the parameters $Q$ and $N$ interwind in the equation of state, the exact expression of the inflection point solution for Eq.~(\ref{cri-con}) is  quite complicated. However, for the special case
of $Q\gg{N}$, the critical point can be approximated as
\begin{equation}\label{cri-ap1}
P_c\approx\frac{1}{96\pi{Q^2}}+\frac{N^2}{216\pi{Q^4}}\;,
V_c\approx8\sqrt{6}\pi{Q}^3+\sqrt{\frac{8}{3}}\pi{Q}N^2\;,
T_c\approx\frac{\sqrt{6}}{18\pi{Q}}+\frac{\sqrt{6}N^2}{72\pi{Q}^3}\;,
\end{equation}
where  the subscript ``c" represents the corresponding  quantity
satisfying the condition for a critical point. As $N$
increases to be large enough, the thermodynamic volume at the inflection point would turn into negative. As result,  the corresponding inflection  point   will become unphysical. Therefore, the additional
constraints $V>0\;, P>0$ and $T>0$  should be considered in Eq.~(\ref{cri-con}). It is
easy to verify that there is an upper bound on  the NUT parameter $N$ for
a fixed eclectic charge $Q$, that is
$N_{b}=\sqrt{3\sqrt{3}-1}Q/2\approx1.0242Q$. If $N>N_{b}$, the
corresponding inflection point  will be located  at where the volume is negative (i.e., $V_c<0$),
which means that there is no longer a physically meaningful critical point or
inflection point.

\begin{figure}[!ht]
\subfigure[]{\label{PV11}
\includegraphics[width=0.45\textwidth]{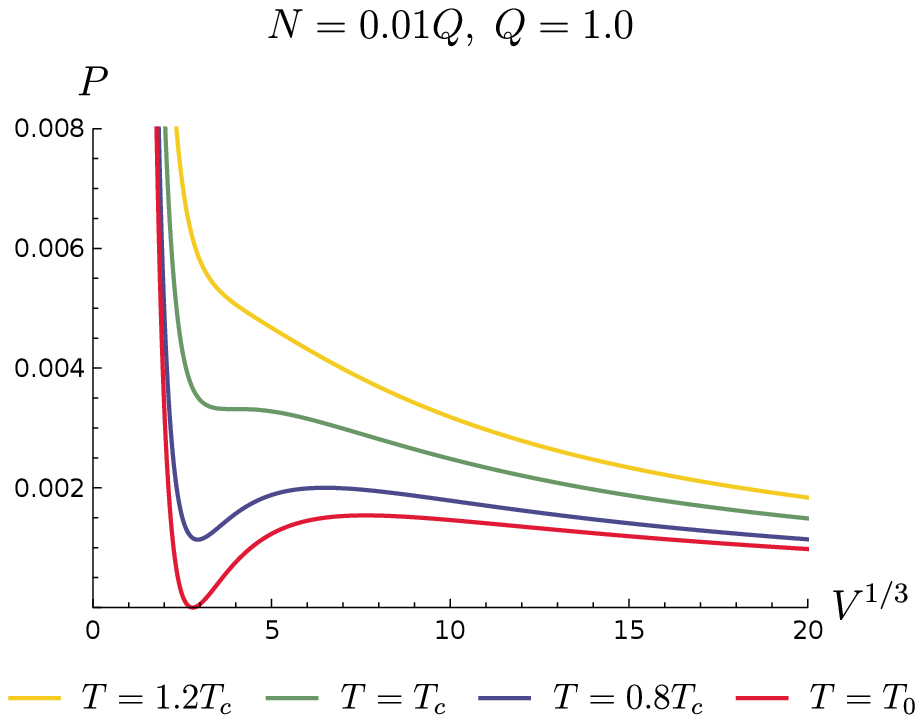}}\qquad\subfigure[]{\label{GT11}
\includegraphics[width=0.45\textwidth]{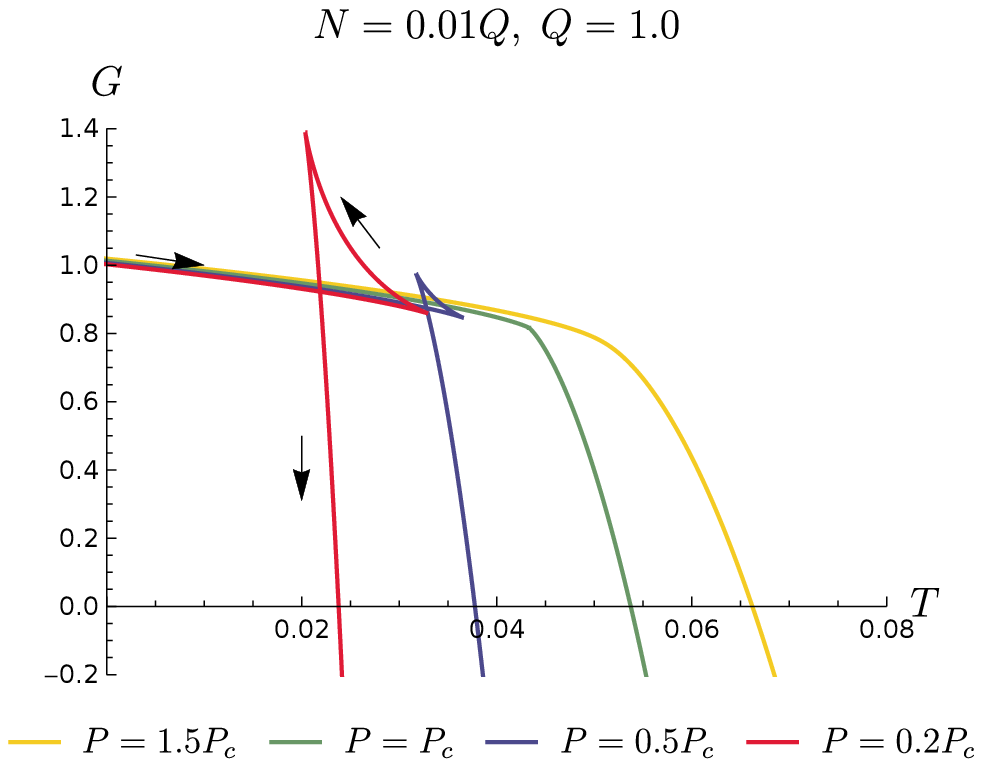}}
\caption{ The plots for the $N$-bihair solution with $Q=1.0, N=0.01Q$. (a) The pressure-volume diagram (the isotherm) is displayed with
$T_c\approx0.0433$ and $T_0\approx0.0306$.  Note that if $T<T_0$, a
part of the isotherm corresponds to a negative pressure. (b) The
behavior of the Gibbs free energy is depicted in the $G$-$T$ plane with
$P_c\approx0.0033$. Its behavior in general is analogous to the situation of
RN-AdS black holes. Here the black arrows indicate the increasing
horizon $r_+$.}\label{PV-GT1}
\end{figure}

According to Eq.~(\ref{state-1}), it is easy to find when the NUT parameter $N$
is very small ($N\ll{N_b}$) the possible phase structure is
analogous to that of the 4-dimensional RN-AdS black holes in the extended
phase space, where the analogous liquid/gas phase transition of the
van der Waals system  occurs~\cite{Mann:2012}.  In order to show
these properties in detail, the corresponding pressure-volume diagram and the
behavior of the Gibbs free energy that is defined by $G=M-TS$ in the extended
thermodynamics for a fixed $Q$ are
illustrated in Fig.(\ref{PV-GT1}). As shown in Fig.(\ref{PV-GT1}), there obviously exists a first order  phase
transition between the small black hole and large black hole for $T<T_c$ or $P<P_c$, and the inflection point obeying
condition~(\ref{cri-con}) represents the critical point of  the phase
transition whose approximate expression is given by
Eq.~(\ref{cri-ap1}).

\begin{figure}[!ht]
\subfigure[]{\label{PV12}
\includegraphics[width=0.45\textwidth]{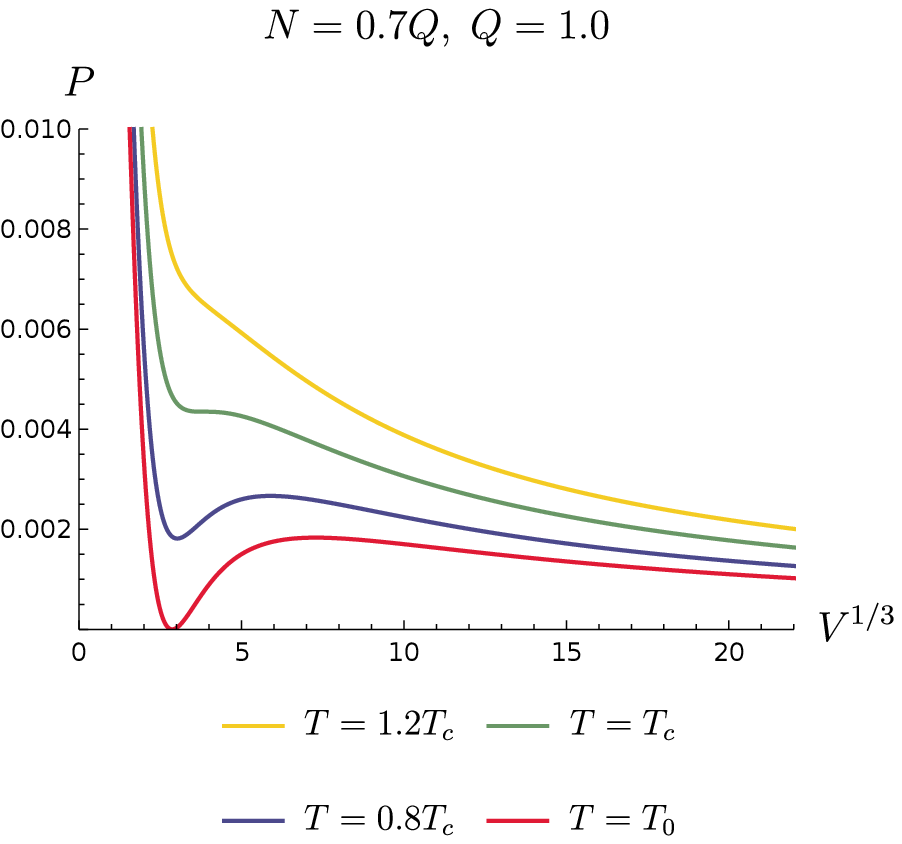}}\qquad\subfigure[]{\label{GT12}
\includegraphics[width=0.45\textwidth]{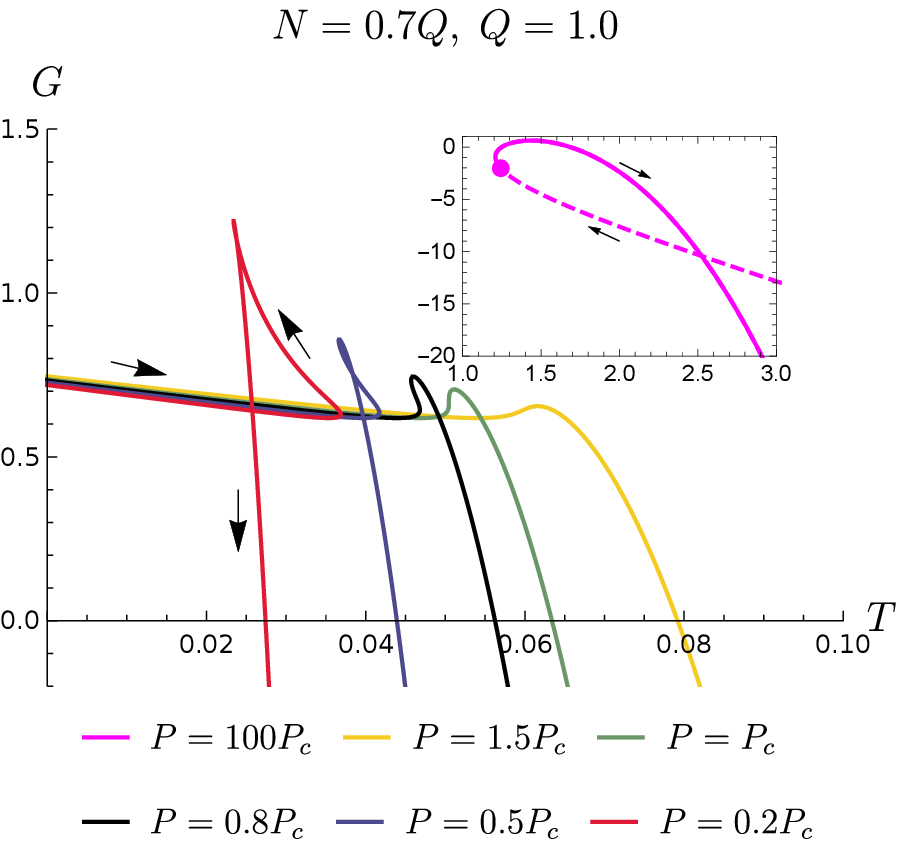}}
\caption{ The plots for the case of the $N$-bihair solution with $Q=1.0,
N=0.7Q$. (a) The pressure-Volume diagram is displayed with $T_c\approx0.0505$
and $T_0\approx0.0337$. (b) The behavior of the Gibbs free energy is
displayed with $P_c\approx0.0044$, where the additional inserted
small  plot  has $P=100P_c$ with the solid circle point
denoting $V=0$.  It should be pointed out that
 the inflection point here is no longer the usual terminal critical point of the first order phase transition due to the presence of a small cusp at $P=P_c$ for the Gibbs  free energy.  Actually,  for small constant
 pressure ($P<0.6289P_c$ in numerical evaluation), there exists  a first order phase transition
between the small and large black hole, and  there seems to be a
zeroth order phase transition for $0.6289P_c<P<P_c$ and
$P>44.50P_c$.  Aside from these regions of $P$,  no  phase
transition occurs. Note the black arrows indicate the increasing
horizon radius $r_+$, and  a zoomed-in  view about the zeroth order phase
transition is displayed in Fig.(\ref{GT-N12zero}).}\label{PV-GT2}
\end{figure}

\begin{figure}[!ht]
\subfigure[]{\label{PV13}
\includegraphics[width=0.45\textwidth]{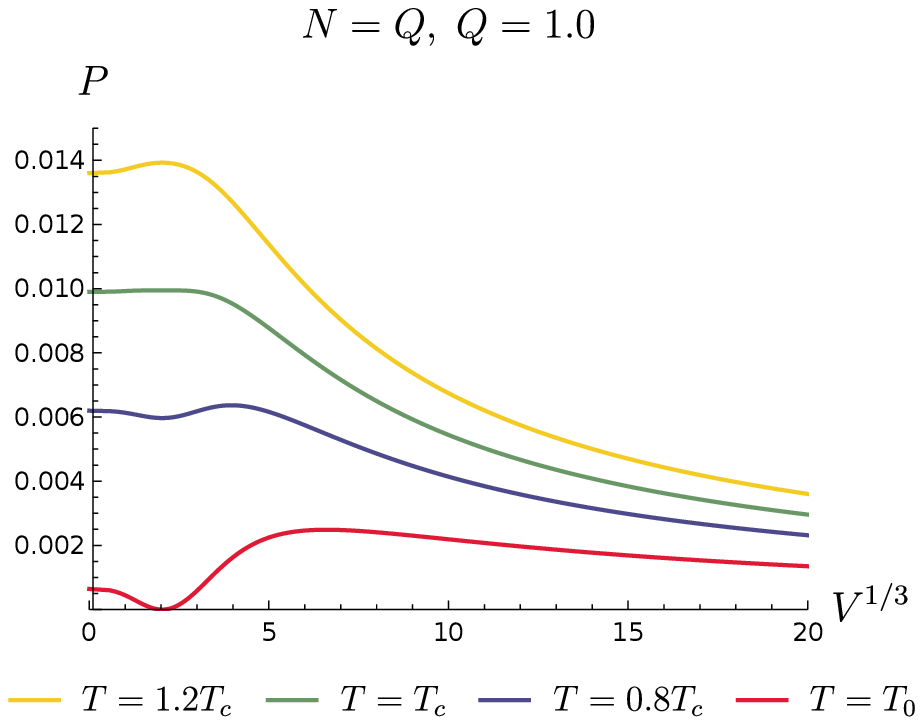}}\qquad\subfigure[]{\label{GT13}
\includegraphics[width=0.45\textwidth]{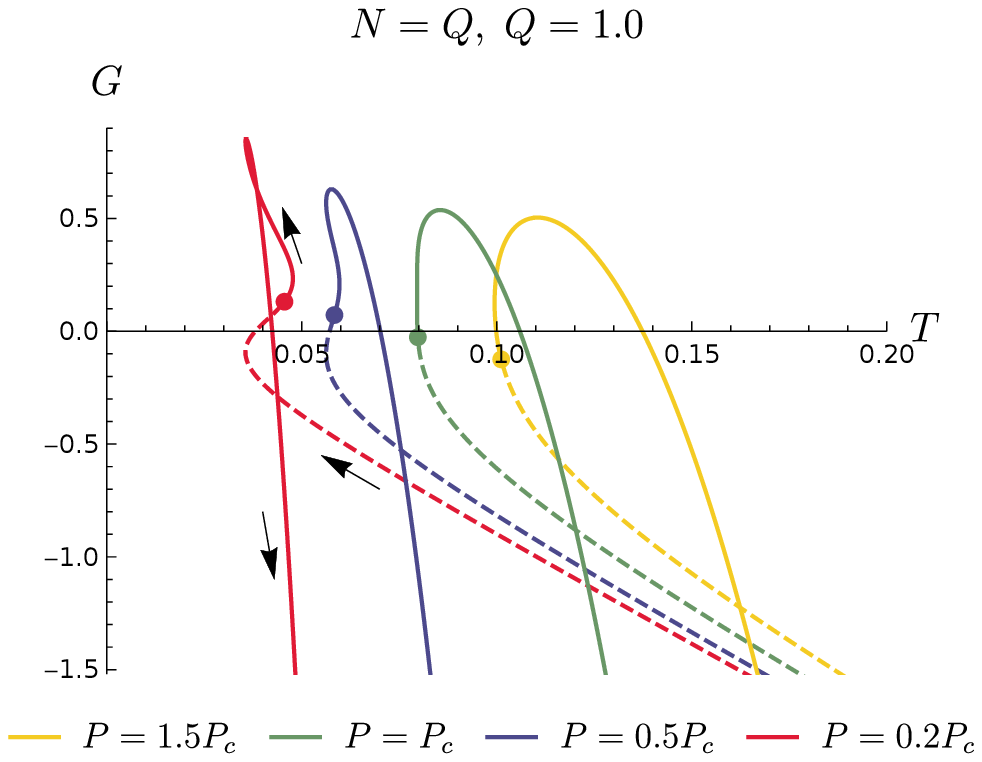}}
\caption{ The plots for the case of the $N$-bihair solution with $Q=1.0,
N=Q$. The isotherms are displayed with  $T_c=1/(4\pi)$ and
$T_0=1/(8\pi)$ in (a). The behavior of the Gibbs free energy is
depicted with $P_c=1/(32\pi)$ in (b), where the  dashed colored
lines and circle points  denote $V<0$ and $V=0$, respectively. The
phase transitions are quite complicated; concretely, for $0<P\leq0.2618P_c$, there exists a first order
phase transition, for $P\in (0.2618P_c,\;0.3030P_c)$,  two first order phase
transitions and one zeroth order phase transition,
for $P\in [0.3030P_c,\;0.3183P_c]$, a first order phase transition and a zeroth order phase
transition, for $P\in (0.3183P_c,\sqrt{3}P_c/2)\cup (\sqrt{3}P_c/2, 0.9306P_c)$,   two zeroth order phase
transitions, and for $P\geq 0.9306P_c$,  only one zeroth order phase transition. The corresponding details
of phase transitions are analyzed  in Fig.~(\ref{GT-N13zero}) as a
supplement. Here the black arrows indicate the increasing horizon $r_+$, }\label{PV-GT3}
\end{figure}

As the value of $N$ increases gradually but still remains smaller than $N_{b}$, the corresponding phase structure
deviates somewhat  from the van der Waals  phase structure. Since the vanishing thermodynamic volume ($V=0$) itself is
more likely  to be located at the phase transition point or inflection point and  plays a crucial role in the behavior of  the phase transition, it is worth exploring firstly the properties of the thermodynamic temperature in the limit of  a vanishing volume.  According to the equation of state~(\ref{state-1}), it is easy to find that  for $N>\sqrt{1+\sqrt{3}}Q/2\approx0.8264Q$, the temperature $T$ at the point of $V=0$ holds  positive  for any  nonnegative pressure, while for $N\leq\sqrt{1+\sqrt{3}}Q/2$, the temperature at $V=0$ would be either positive or negative,  strongly depending on the magnitude  of pressure $P$.  After some simple algebraic manipulations, such  pressure can be straightforwardly obtained
\begin{equation}
P_0=\frac{Q^2-2(\sqrt{3}-1)N^2}{64\pi(2-\sqrt{3})N^4}\;.
\end{equation}
 If $P>P_0$,  the temperature at
$V=0$ is positive for $N\leq\sqrt{1+\sqrt{3}}Q/2$; however, for $P<P_0$,
the temperature at $V=0$ will be negative for
$N\leq\sqrt{1+\sqrt{3}}Q/2$. The latter case implies that $V=0$ will
make the temperature  be of no physical sense.  These properties can
be straightforwardly captured in  Figs.~(\ref{PV-GT2})
and~(\ref{PV-GT3}) that follow. Indeed, if we want to explore the interesting
role  played by $N$ and the vanishingly small $V$ on the
thermodynamical phase transition behavior, the non-negativeness of
the  thermodynamical  temperature is a necessary constraint.
Therefore, we can examine, without loss generality, the
corresponding behavior of phase transitions for the $N$-bihair solution
via appropriately  choosing  certain values of $N$
based on the above discussions.

\begin{figure}[!ht]
\subfigure[]{\label{GT-N12zero11}
\includegraphics[width=0.45\textwidth]{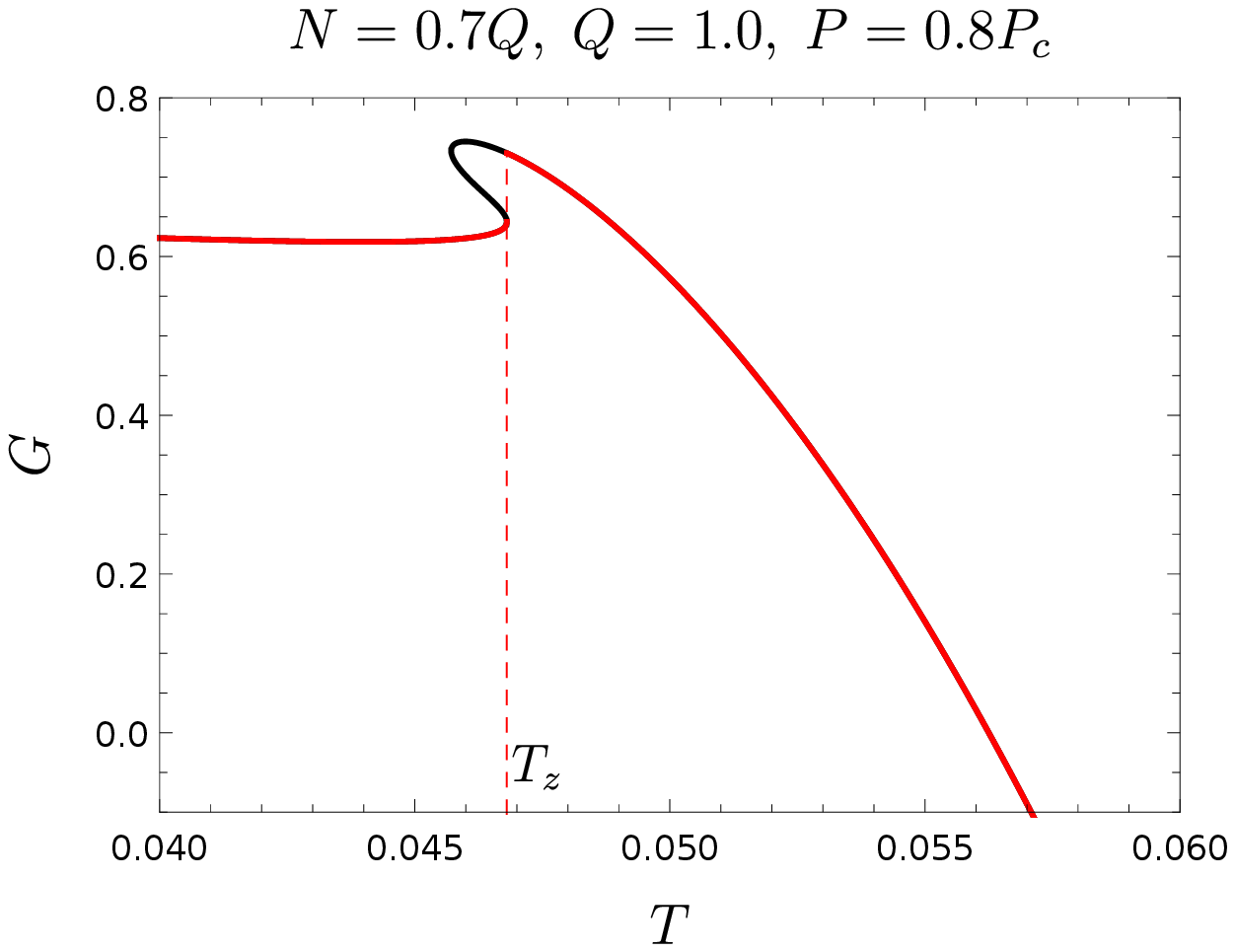}}\qquad\subfigure[]{\label{GT-N12zero22}
\includegraphics[width=0.45\textwidth]{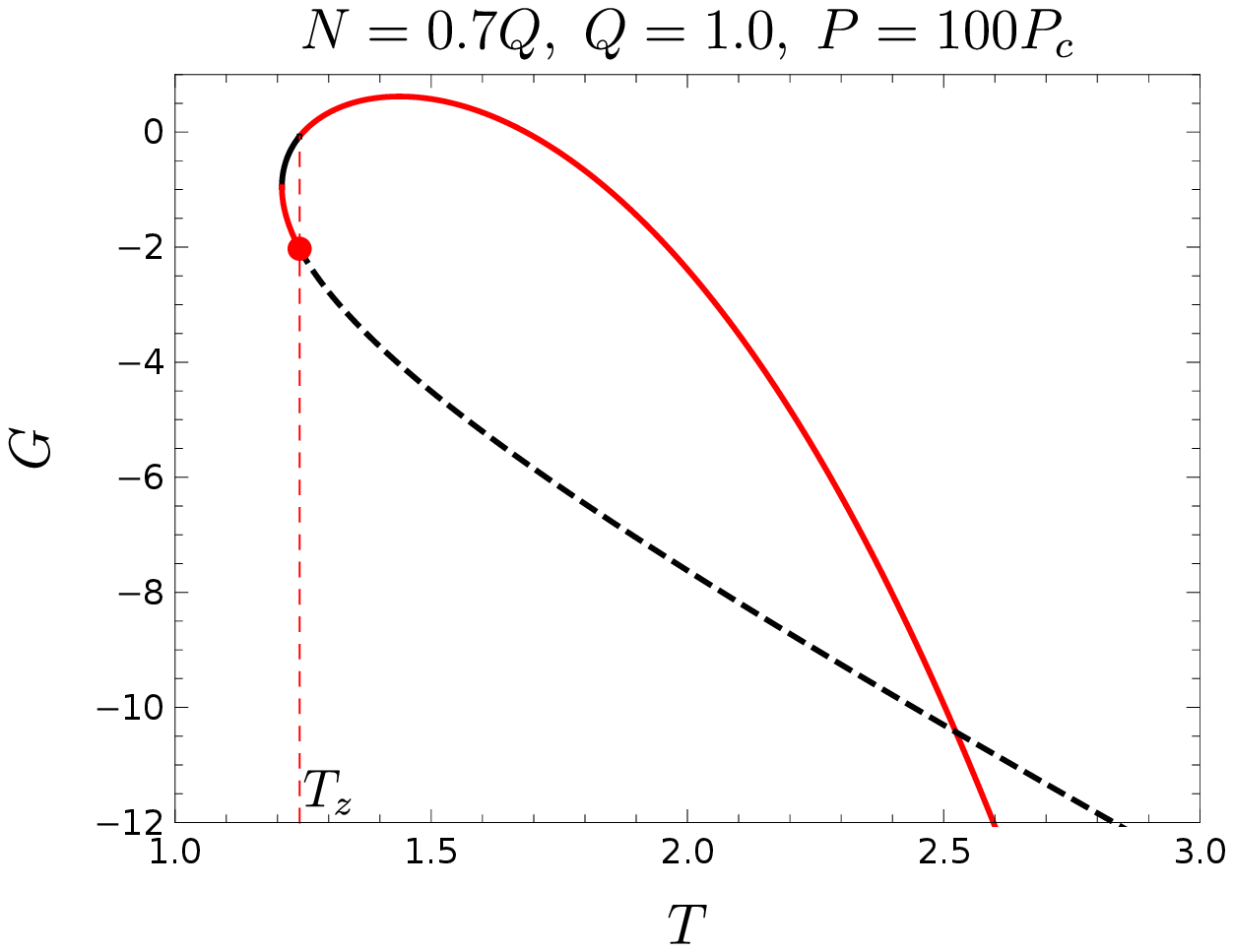}}
\caption{ A zoomed view of the zeroth order phase transition in the case of
$N$-bihair solution with $Q=1.0, N=0.7Q, P_c\approx0.0044$ for (a) $P=0.8P_c$ and for
(b) $P=100P_c$. The red solid lines indicate the global minimal
branch of the Gibbs free energy, while the black solid lines denote
the non-minimal branch. The black dashed line and the circle point
in right plot stand for $V<0$ and $V=0$, respectively. Obviously,
there is a jump for the global minimum of Gibbs free energy at
temperature $T_z$, which means the occurrence of zeroth order
phase transition between the small and large black
hole.}\label{GT-N12zero}
\end{figure}

In Fig.~(\ref{PV-GT2}), with $N=0.7Q<\sqrt{1+\sqrt{3}}Q/2$,  we find  as we expect that only if the pressure is large enough,  can  the point
$V=0$  occur as an obstruction of the Gibbs free energy in the $G$-$T$
diagram. Interestingly, there, aside from the first order phase
transition for $P<0.6289P_c$, exists a zeroth order phase transition
at a certain large constant pressure (e.g., $P=100P_c>P_0$). The
zeroth order phase transition is not a real phase transition, but it
is discovered that the global minimum of the Gibbs free energy is
discontinuous in the $G$-$T$ diagram~\cite{Gunasekaran:2012}.   The
details of the zeroth order phase transition are depicted in
Fig.~(\ref{GT-N12zero}). Therefore, one may  conclude that the
point $V=0$ located on the minimal branch of the Gibbs free energy
for the $N$-bihair solution  may have important effects on the possible
phase transitions. Besides these, it should be pointed out the inflection point here is no longer the usual thermodynamic terminal critical point of the first order phase transition.

\begin{figure}[!htp]
\subfigure[]{\label{GT-N13zero11}
\includegraphics[width=0.325\textwidth]{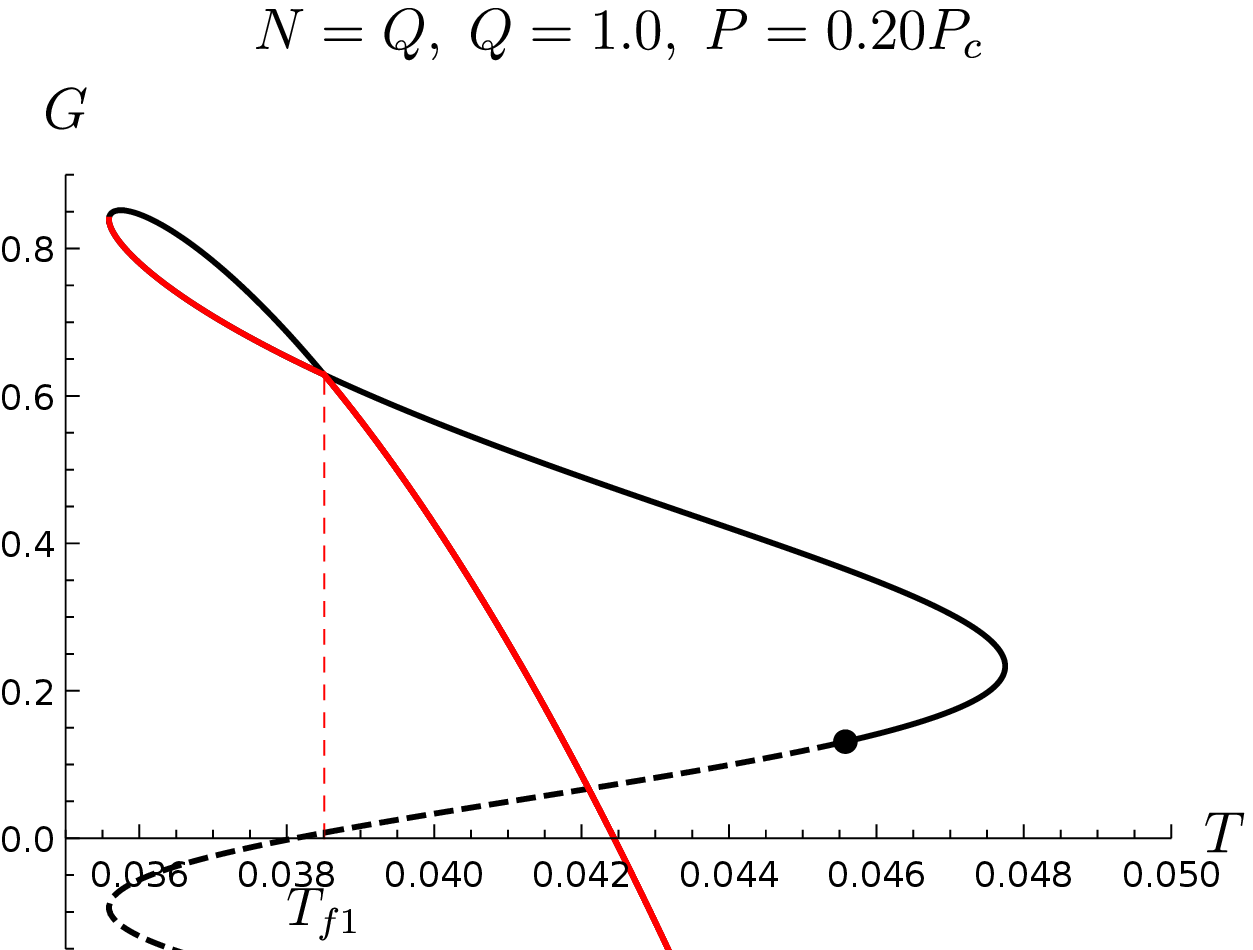}}\;\subfigure[]{\label{GT-N13zero22}
\includegraphics[width=0.325\textwidth]{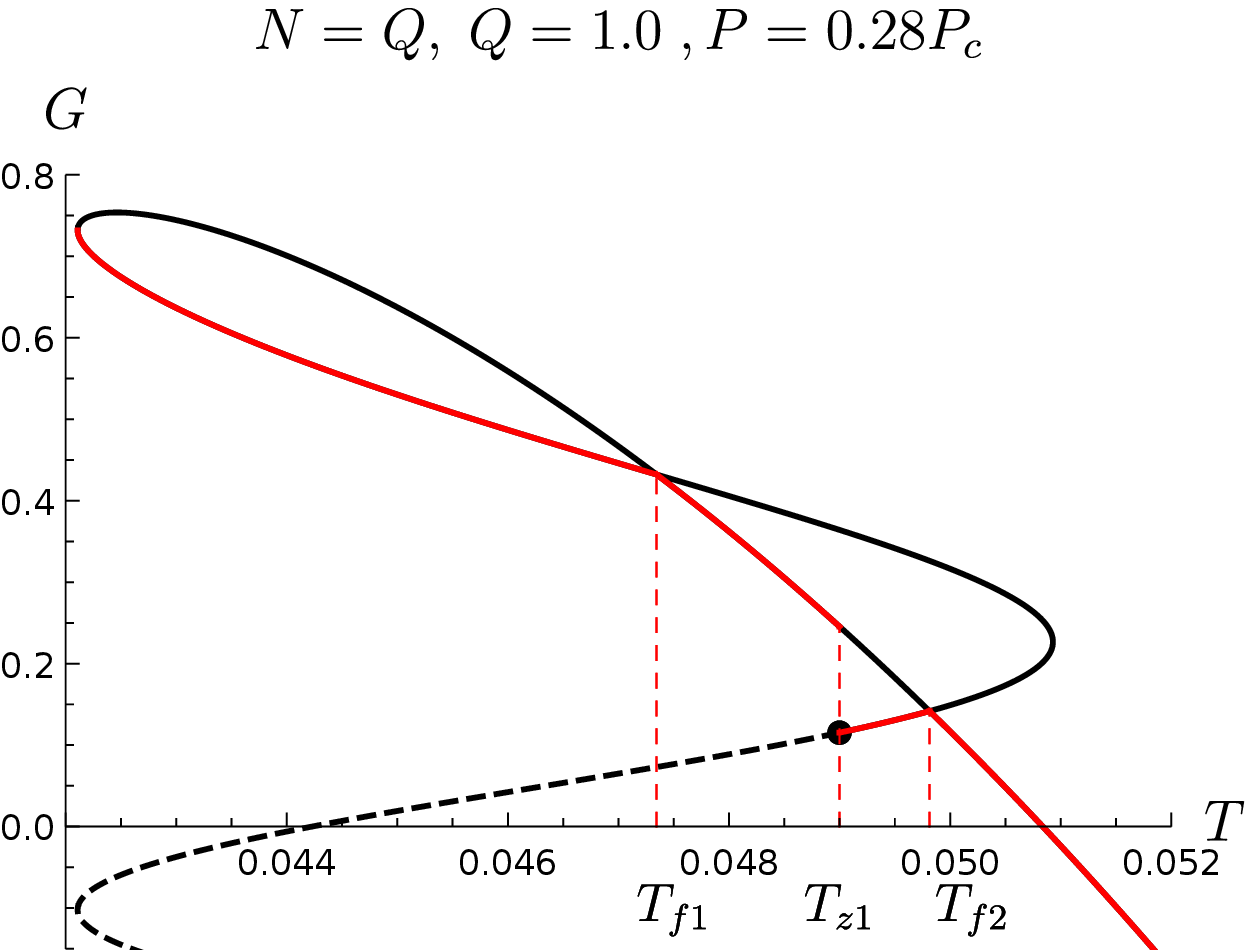}}\;\subfigure[]{\label{GT-N13zero33}
\includegraphics[width=0.325\textwidth]{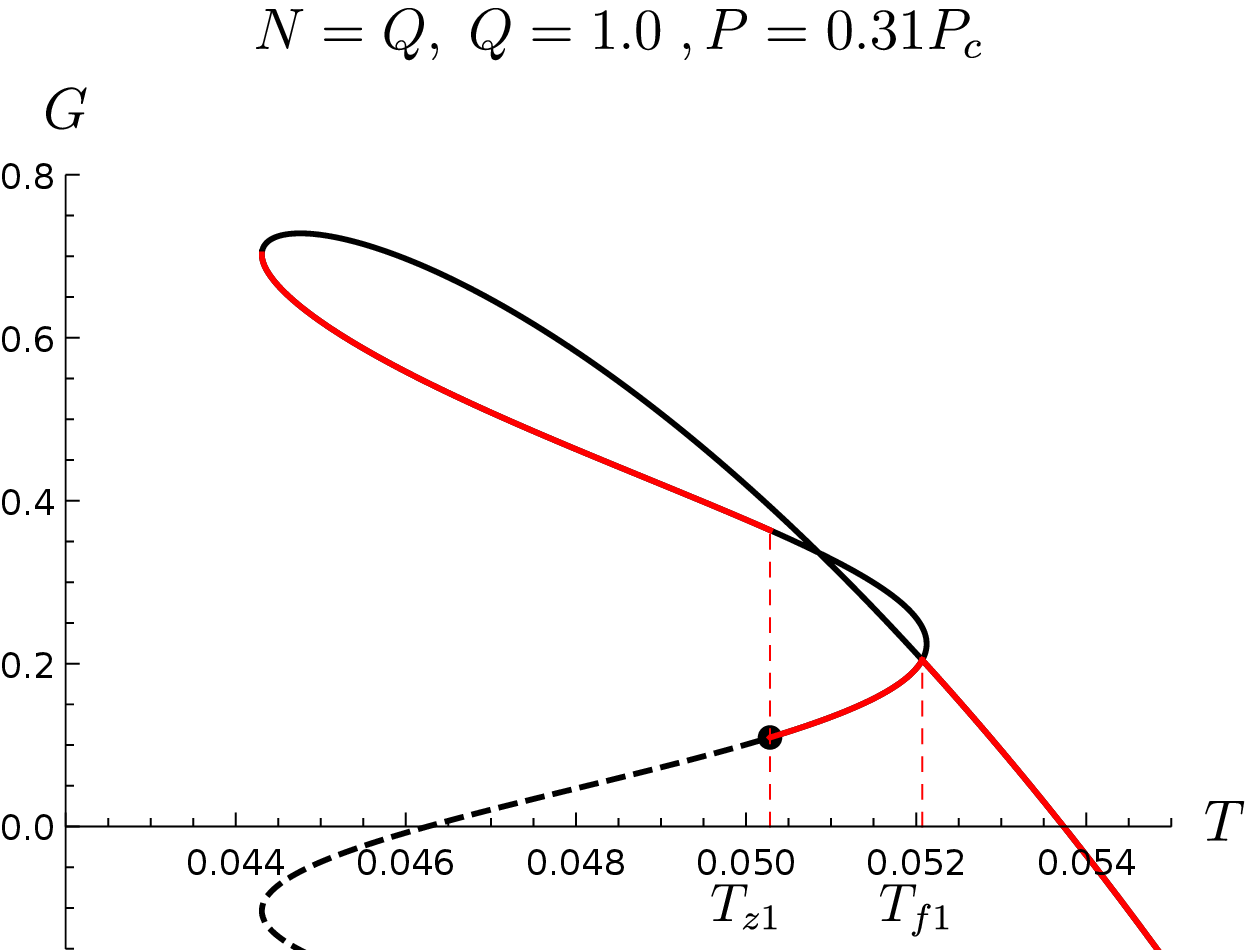}}\;\\\subfigure[]{\label{GT-N13zero44}
\includegraphics[width=0.325\textwidth]{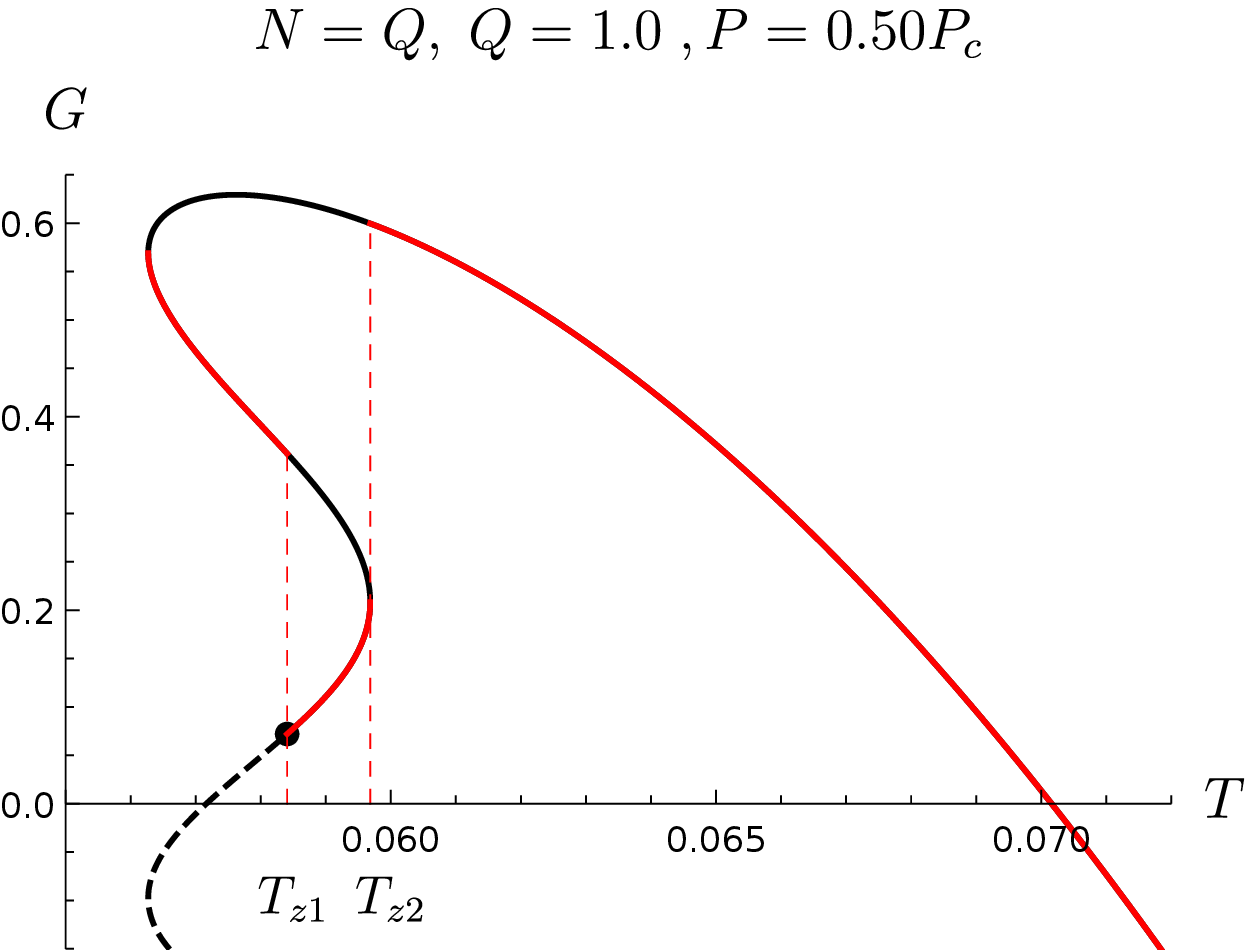}}\;\subfigure[]{\label{GT-N13zero55}
\includegraphics[width=0.325\textwidth]{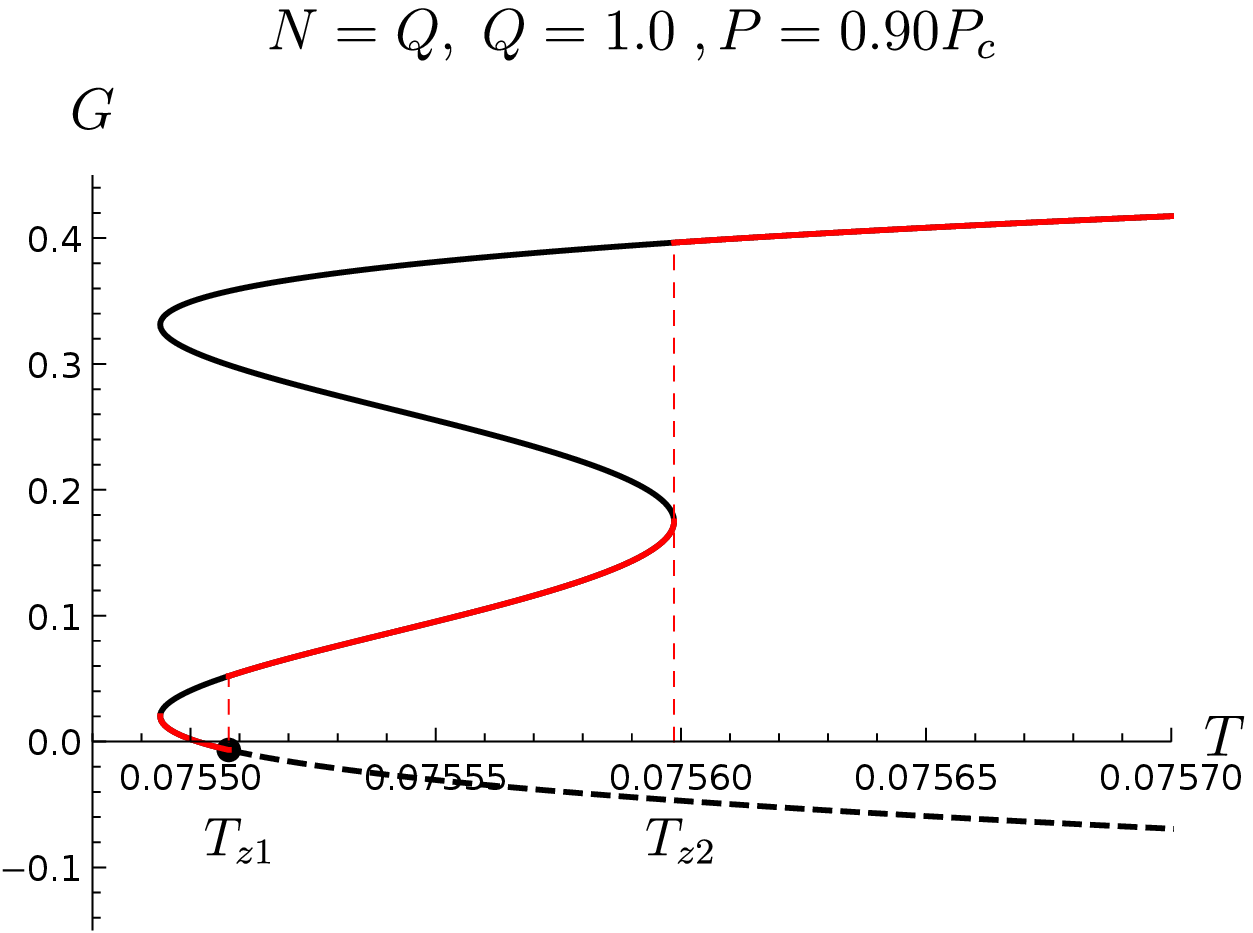}}\;\subfigure[]{\label{GT-N13zero66}
\includegraphics[width=0.325\textwidth]{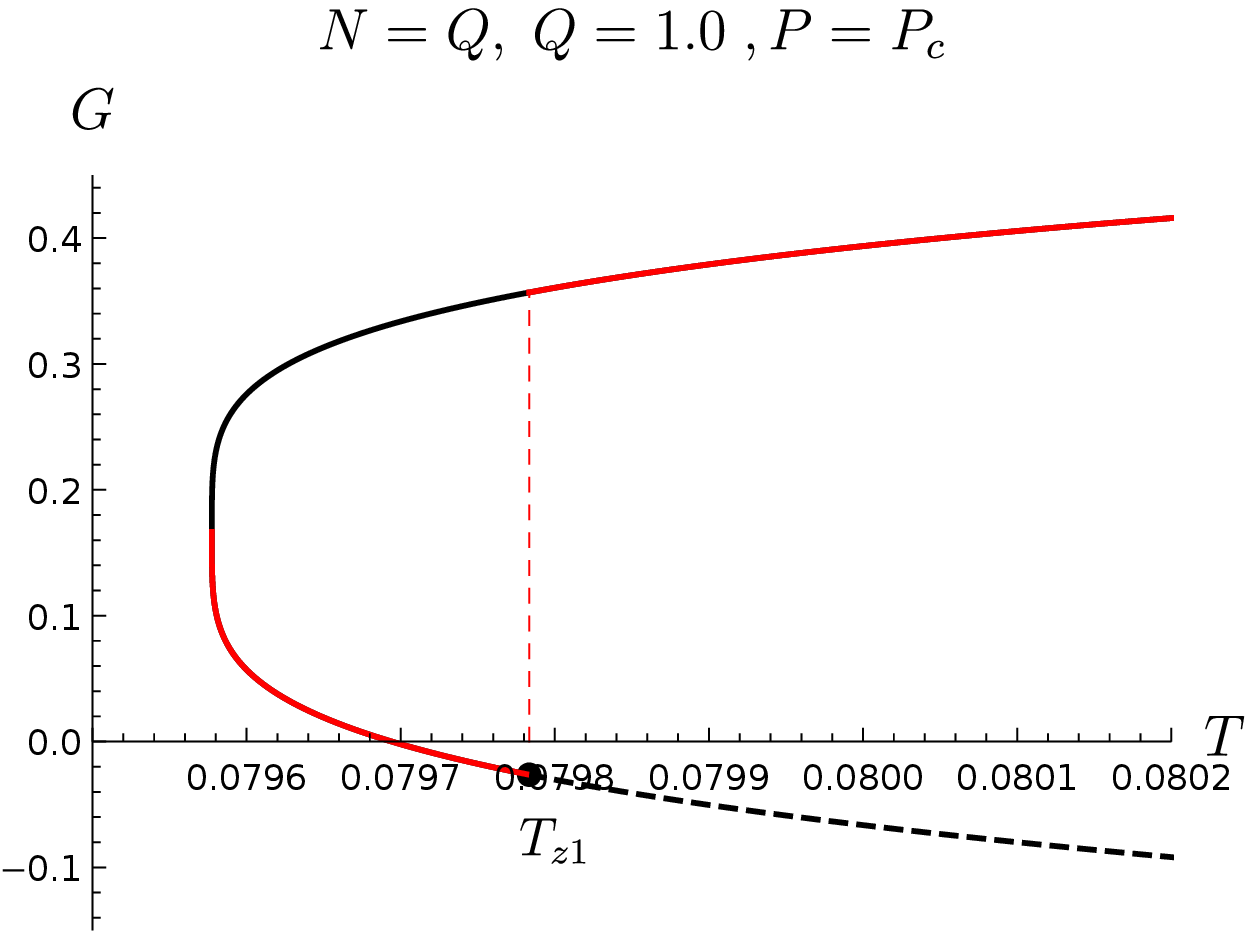}}
\caption{ Some zoomed-in views of the possible  phase transitions for the
$N$-bihair solution with $Q=1.0, N=Q$ and $P_c=1/(32\pi)$. In all
plots, the red solid lines highlight the global minimal branches of
the Gibbs free energy, while the black solid lines denote the
non-minimal branches with the black dashed lines and the circle
points indicating $V<0$ and $V=0$, respectively. Here, $T_{z1},
T_{z2}$ indicate the corresponding temperatures of the zeroth order
phase transition and $T_{f1},T_{f2}$ correspond the temperatures of
the first order phase transition. More interestingly, for the case
of $P=0.28P_c$, there exists a reentrant-like phase transition
behavior: the large/small/large/medium black hole phase transition
as the temperature continuously decreases.}\label{GT-N13zero}
\end{figure}

As shown in Fig.~(\ref{PV-GT3}),  with $N=Q$ satisfying
$\sqrt{1+\sqrt{3}}Q/2<N<N_b$, the phase structure becomes quite rich
in comparison with the case of $N<\sqrt{1+\sqrt{3}}Q/2$. There still
exist possible  zeroth  and first order phase transitions in the
$G$-$T$ diagram, and it is highly possible that the points of $V=0$
just correspond to the zeroth order phase transition points as long as the
constant pressure $P$ is not too small. Moreover,  as shown in
Fig.~(\ref{GT-N13zero}), where  the possible phase transition points
are marked by vertical red dashed lines, besides the large/small black
hole phase transition, there exist some complicated phase
transitions, e.g.,  the large/small/large/medium black hole phase
transition, the large/small/medium black hole phase transition and the large/medium/small black hole phase transition. Here,
the inflection point in this case is again no longer the usual terminal critical
point of the first order phase transition, and the behavior
of the phase transition is quite different from that of the van der Waals
phase transition.

\begin{figure}[!ht]
\subfigure[]{\label{PV14}
\includegraphics[width=0.45\textwidth]{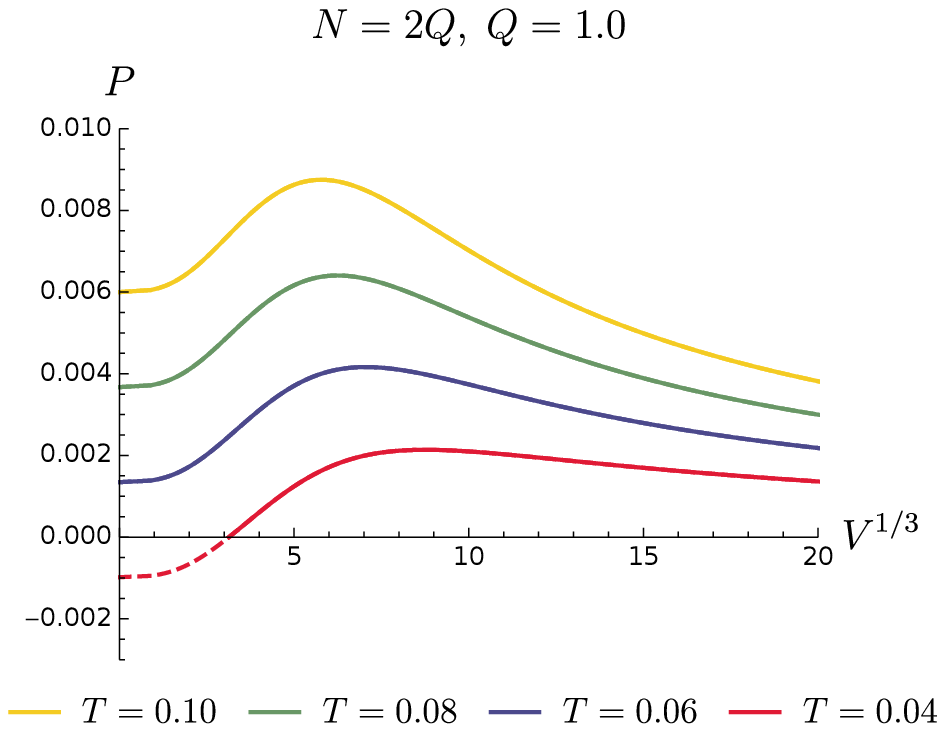}}\qquad\subfigure[]{\label{GT14}
\includegraphics[width=0.45\textwidth]{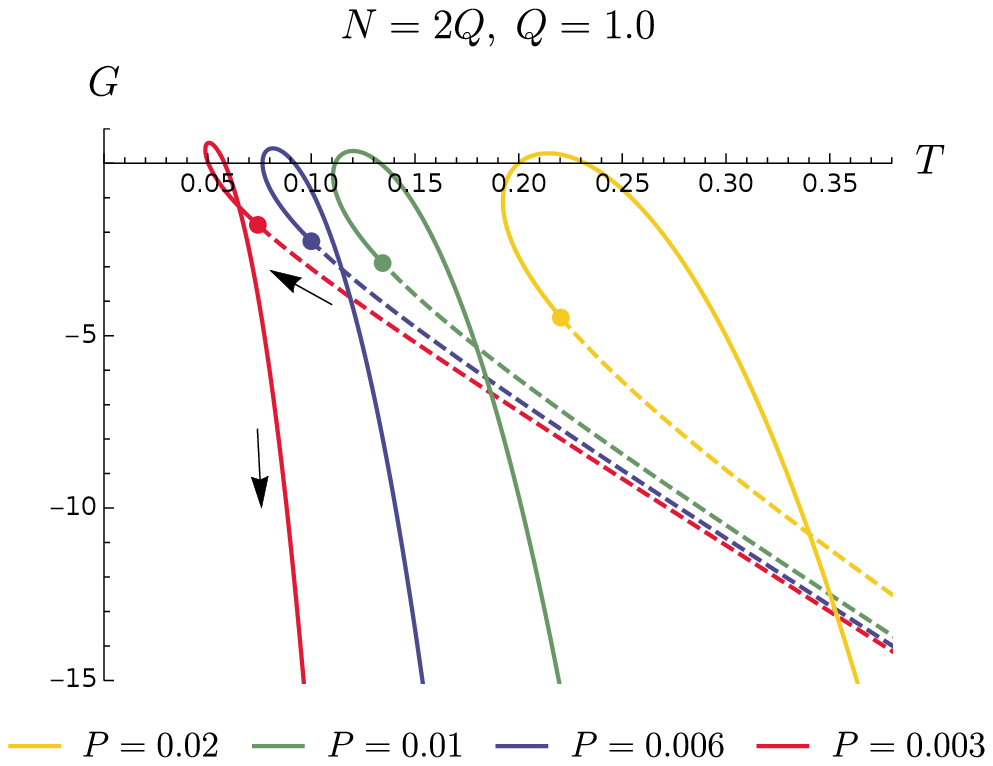}}
\caption{ The plots for the case of $N$-bihair solution with $Q=1.0,
N=2Q$. The isotherms  and  Gibbs free energy  are displayed in (a)
and (b), respectively.  Note that the dashed lines correspond to negative pressure in (a) and negative volume in (b), respectively.
For $P\leq0.0041$, there is a first order phase transition, while for $P>0.0041$, the points of $V=0$, denoted by circle points, indicate the zeroth order phase transitions. The black arrows indicate the increasing horizon $r_+$.}\label{PV-GT4}
\end{figure}

As for a large $N$,  we can set $N=2Q>N_b$ without loss of generality, then there is no longer physical inflection point for the thermodynamic
 system (see Fig.~(\ref{PV-GT4})).  And if the pressure is large enough,  all the points of $V=0$ in the $G$-$T$ diagram correspond to the points of
the zeroth order phase transition between the small and large black
hole.

\subsection{$N$-trihair solution}
Let us now turn to  the $N$-trihair solution.
Similarly,  the equation of state can be written as
\begin{equation}\label{state-2}
P=\frac{r_+T}{2(N^2+r_+^2)}-\frac{1}{8\pi(N^2+r_+^2)}+\frac{Q^2}{8\pi(N^2+r_+^2)^2}\;,
\widetilde{V}=\frac{4\pi r_+(r_+^2+3N^2)}{3}\;.
\end{equation}
The critical point (inflection point) can still be obtained  by
solving the equation
\begin{equation}\label{cri-con-2}
\frac{\partial P}{\partial \widetilde{V}}=0\;,~\frac{\partial^2
P}{\partial \widetilde{V}^2}=0\;.
\end{equation}
To make the inflection have physical sense,  constraints
$\widetilde{V}>0\;, P>0$ and $T>0 $  should also be imposed on
Eq.~(\ref{cri-con-2}).

After some straightforward algebraical manipulations, we find that there
also exists  an upper bound on $N$ (denoted by $N_t$ ), that is
$N_t=3Q/2\sqrt{2}\approx1.0607Q$ which is slightly larger than
the previous one for the $N$-bihair situation. It should be
emphasized that for $N>N_t$ the inflection point would become
unphysical, since a large value of $N$ would render the pressure at
the inflection point negative rather than the volume.  For a
special case of $N\ll{Q}$, the inflection  point can be
simplified to
\begin{equation}\label{cri-ap2}
P_c\approx\frac{1}{96\pi{Q^2}}+\frac{N^2}{216\pi{Q^4}}\;,
\widetilde{V}_c\approx8\sqrt{6}\pi{Q}^3-2\sqrt{6}\pi{Q}N^2\;,
T_c\approx\frac{\sqrt{6}}{18\pi{Q}}+\frac{\sqrt{6}N^2}{72\pi{Q}^3}\;.
\end{equation}

\begin{figure}[!htp]
\subfigure[]{\label{PV23}
\includegraphics[width=0.45\textwidth]{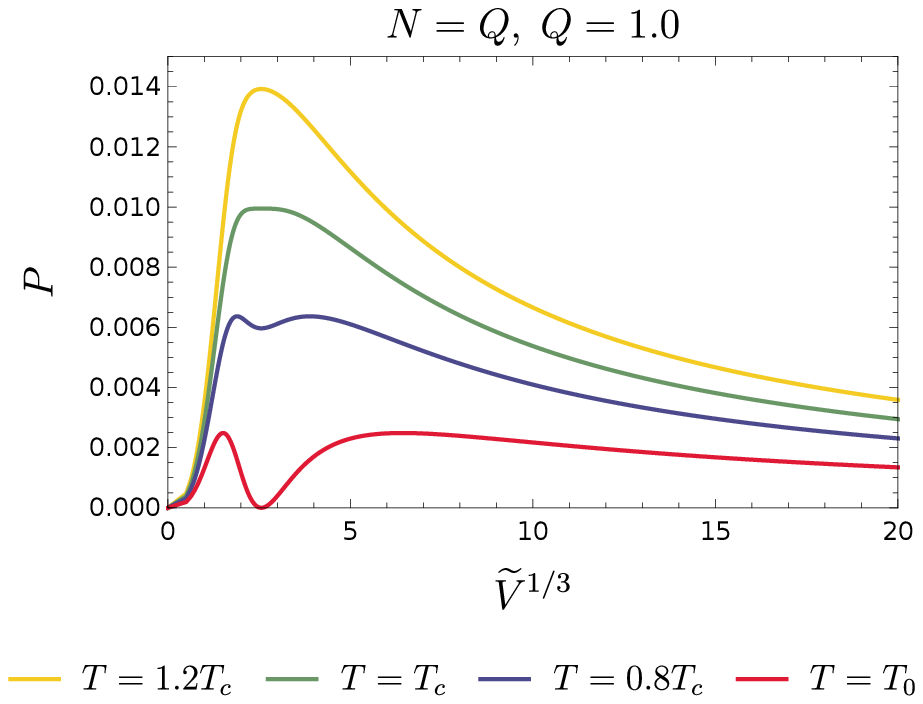}}\qquad\subfigure[]{\label{PV24}
\includegraphics[width=0.45\textwidth]{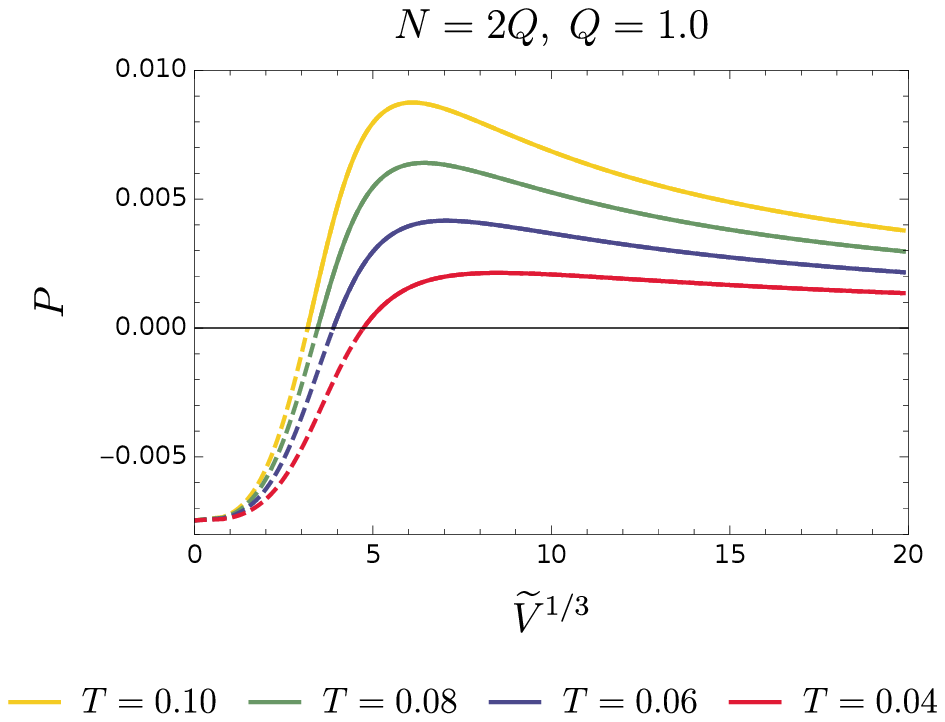}}
\caption{ The pressure-volume diagram for the $N$-trihair solution at various
temperature with fixed $Q=1.0, N=Q$ in (a) and $Q=1.0, N=2Q$ in (b).
 Here, $T_c=1/(4 \pi)$, $T_0=1/(8\pi)$ and the dashed lines represent the negative pressure.}\label{PV2324}
\end{figure}

 According to the equation of state~(\ref{state-2}) and Eq.~(\ref{cri-ap2}), it is
easy to deduce that for small $N$ the phase structure of the $N$-trihair
solution is quite analogous to that of the $N$-bihair solution. If $N$ is not
too small, according to the  foregoing discussion, the  influence of
$N$ on the phase transitions  would become important.  Without loss generality, we could choose $N=\{Q, 2Q\}$ for convenience in comparison with the $N$-bihair solution.   In
Fig.~(\ref{PV2324}), the isotherms for the  $N$-trihair solution with
$N=Q$ and $N=2Q$ are plotted, respectively.  We can see that the
pressure $P$  is  an increasing function of the  volume
$\widetilde{V}$ in the regime of small volume, especially in
comparison with the $N$-bihair solution at $N=Q$ (Fig.~(\ref{PV13})), and such characters of
increasing function are quite apparent. One may analyze this  property
from a pure mathematical point of view. Differentiate the state equations~(\ref{state-1})(\ref{state-2}), we
have
\begin{equation}\label{Nb21}
\Big(\frac{\partial{P}}{\partial{V}}\Big)_{T}=\Big(\frac{\partial{P}}{\partial{r_+}}\Big)_{T}\frac{\partial{r_+}}{\partial{V}}
=-\frac{3}{16\pi^2}\frac{(N^2-2Q^2)r_++r_+^3+2N^4T\pi-2\pi{T}r_+^4}{(N^2+r_+^2)(3N^6-21N^4r_+^2-11N^2r_+^4-3r_+^6)}\;
\end{equation}
and
\begin{equation}\label{Nb22}
\Big(\frac{\partial{P}}{\partial{\widetilde{V}}}\Big)_{T}
=\frac{r_+(N^2-2Q^2+r_+^2)}{16\pi^2(N^2+r_+^2)^4}+\frac{(N^2-r_+^2)T}{8\pi(N^2+r_+^2)^3}\;.
\end{equation}
For the $N$-bihair solution, whether $\big({\partial{P}}/{\partial{V}}\big)_{T}$ is
negative or positive near the point $V=0$ (i.e., $r_+=\sqrt{2\sqrt{3}-3}N$) should be dependent on both the NUT
parameter and the temperature, while for the $N$-trihair solution,  for vanishingly small $\widetilde{V}$ ($r_+\rightarrow0$),
Eq.~(\ref{Nb22}) can be approximated as
\begin{equation}
\Big(\frac{\partial{P}}{\partial{\widetilde{V}}}\Big)_{T}\approx\frac{T}{8N^4\pi}+\frac{r_+(N^2-2Q^2)}{16N^8\pi^2}\;.
\end{equation}
 Thus, the thermodynamic pressure $P$ at a constant temperature is an
increasing function of $\widetilde{V}$ in the regime of small volume as long as $N$ is not vanishingly small.

\begin{figure}[!htp]
\includegraphics[width=0.7\textwidth]{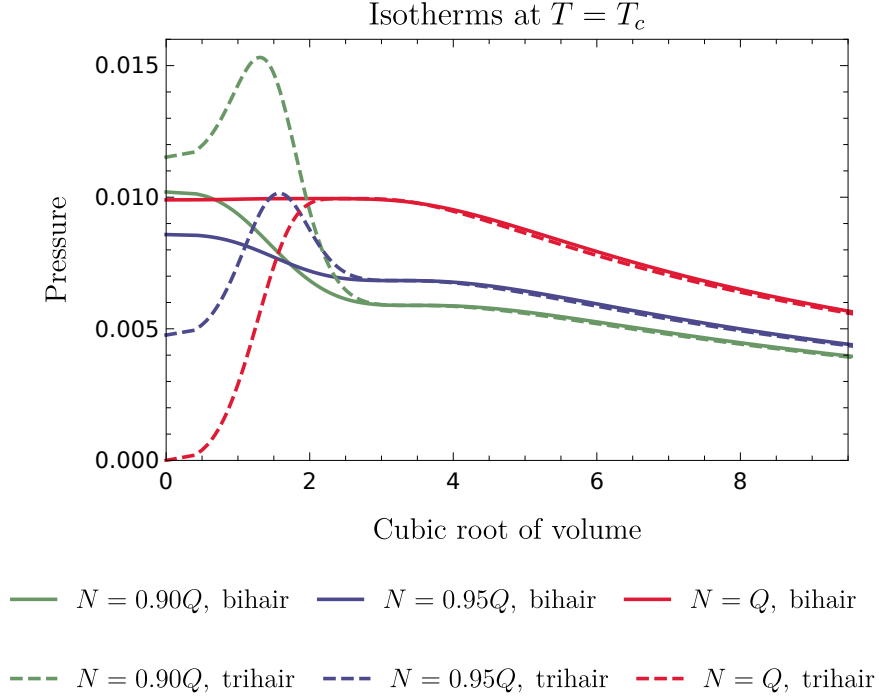}
\caption{ The isotherms at the temperature $T=T_c$ for both
$N$-bihair and $N$-trihair solution with fixed $Q=1.0$.
}\label{Tcall}
\end{figure}

 For a cross-comparison of  the pressure-volume diagram between the $N$-bihair  and  $N$-trihair
solutions, the isotherms at temperature $T=T_c$
for  both the $N$-bihair  and $N$-trihair situation are plotted in Fig.(\ref{Tcall})
as a supplement to analyze above property.

\begin{figure}[!ht]
\subfigure[]{\label{GT23}
\includegraphics[width=0.45\textwidth]{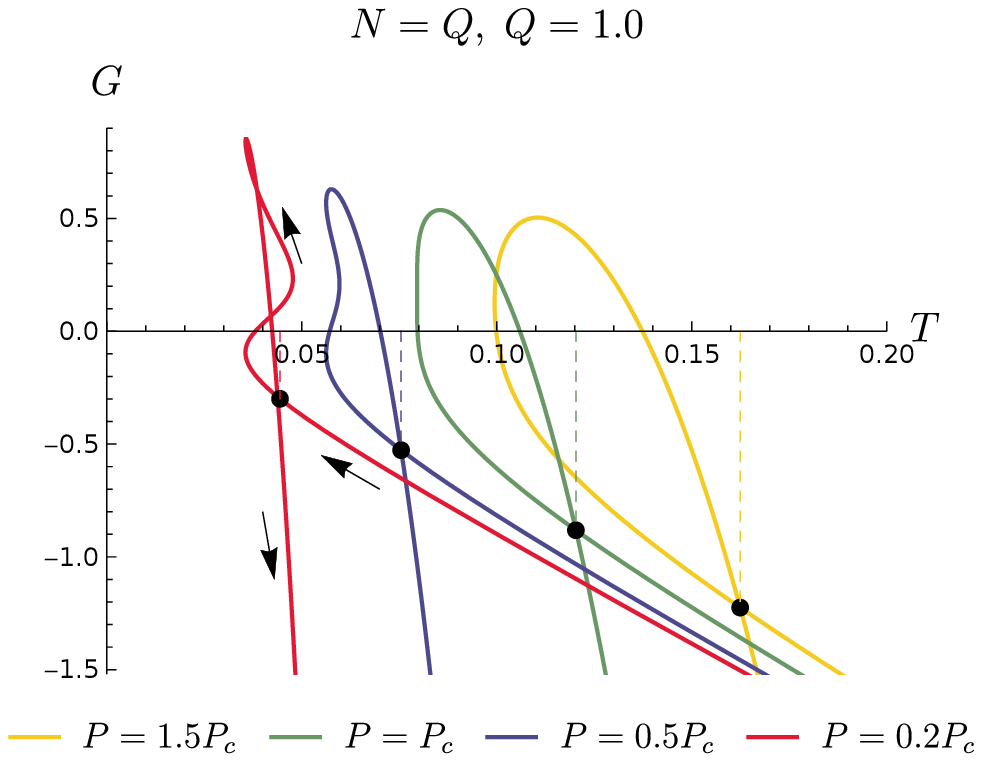}}\qquad\subfigure[]{\label{GT24}
\includegraphics[width=0.45\textwidth]{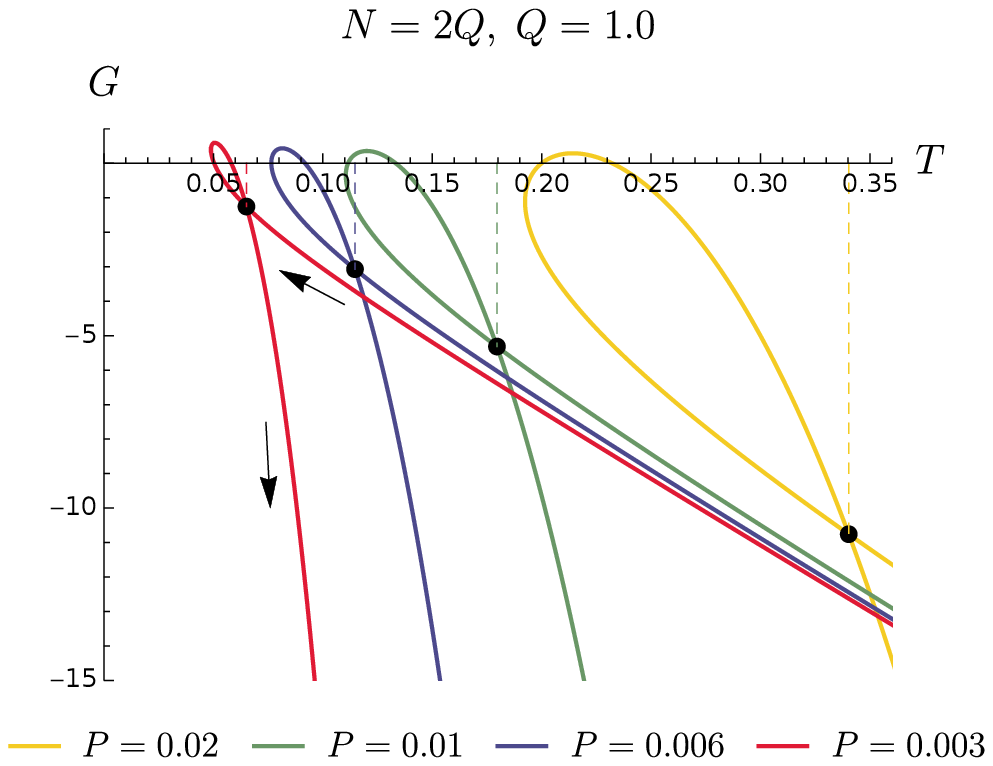}}
\caption{ The Gibbs free energy is plotted as a function of
temperature in the $G$-$T$ plane for  the $N$-trihair solution. Here, we set
$Q=1.0, N=Q$ in (a) and $Q=1.0, N=2Q$ in (b). The vertical dashed
colored lines or the black circle dots indicate the corresponding
temperatures where the first order phase transition would occur at a
certain constant pressure. The black arrows indicate the increasing
horizon radius $r_+$.}\label{GT2324}
\end{figure}

In Fig.~(\ref{GT2324}), the  behavior of  the Gibbs free energy for the $N$-trihair solution with $N=\{Q, 2Q\}$ is depicted
 in detail. It is not difficult to find  that all the curves  are  identical
to those for the $N$-bihair solution except for that the dashed curves of the Gibbs free energy correspond to the constraint of a positive
volume. Here, it should be emphasized that as long as $r_+>0$,
$\widetilde {V}>0$ will hold. Therefore, the constraint of
$\widetilde {V}>0$  is trivial and independent of $N$, which leads
to that the first order phase transition occurs for the $N$-trihair
solution rather than the zeroth phase transition. The detailed phase
transition points are indicated by vertical dashed red lines.
However, the analogy to  the van der Waals  liquid-gas phase
transition has been  broken  by  the presence of a not too small $N$.
The  inflection points, if exist,  do not indicate the usual
critical points of thermodynamic phase transitions any more.

\section{Conclusion}

In the extended thermodynamic phase space, we have
studied the thermodynamical phase structure of the 4-dimensional  Lorentzian
RN-Taub-NUT-AdS spacetime, including the possible phase transitions,
critical points and analogy  to the van der Waals fluid.  Armed
with the cognition that the NUT parameter may possess multiple
physical characters and  the consistent
thermodynamical  first law  formulated in Ref.~\cite{Wu:2019}, we analyze the
influence of the NUT parameter on the possible phase transitions of
such NUT-type spacetime in detail.

When the NUT parameter $N$ is interpreted as a thermodynamic
bihair, we find that there is an upper bound on the value of $N$ for
fixed electric charge $Q$, that is $N_b=\sqrt{3\sqrt{3}-1}Q/2$,
beyond which the corresponding physical inflection point or critical
point will not exist. If the NUT parameter $N$ is very small
($N\ll{N_b}$), there seems to be a first order phase transition
between  the small and large black hole, which is quite analogous to
the van der Waals  phase transition between liquid and gas. As $N$
increases gradually, the analogy to the van der Waals system is
broken somewhat. The zeroth order phase transition between the small
and large black hole may occur, and the inflection point obtained by
solving the condition for a critical point is not the usual
thermodynamic critical point. Especially for $N=Q$, the constraint
of a positive volume may give rise to a zeroth order phase
transition.  As a result, there exists rich phase structures and
more complicated reentrant-like phase transitions. With $N$
increasing beyond the upper bound $N_{b}$, it is found that the
first order phase transition and the inflection point will disappear
completely for a large enough positive pressure. Instead,  only the
zeroth order phase transition is likely to happen.

Regarding the $N$-trihair solution, one may also find a upper bound,
$N_t=3Q/2{\sqrt{2}}$, which is a little larger than the upper bound
$N_b$ for  the $N$-bihair solution. For small $N$($N<Q$), the
small/large black hole phase transition, similar to the $N$-bihair
solution,  will appear like the van der Waals liquid/gas phase
transition. As $N$ increases to be comparable to $Q$,  in contrast
to the situation of the $N$-bihair solution, the phase structure then
will become  relatively simple, and it seems that only the first order phase
transition between the small and large black hole occurs since  the
mere constraint of a positive volume cannot give rise to the zeroth
order phase transition for the $N$-trihair solution. Particularly,
when $N$ goes beyond its upper bound $N_t$, there is neither the
zeroth order phase transition nor the physical inflection point or
critical point. Besides these, for not too small $N$, it is
obvious that the pressure along the isotherm  for the $N$-trihair
solution  behaves as an increasing function of $\widetilde{V}$ in
the regime of small volume. Therefore, one may conclude that not
only the magnitude  but also the thermodynamic multihair characters
of  the NUT parameter can influence the  phase structure and the
phase transition of the 4-dimensional RN-Taub-NUT-AdS spacetime. The
thermodynamical analogy to  the  van der Waals system is valid only
in the case of a vanishingly small NUT parameter.

\begin{acknowledgments}
 We would like to thank  Di Wu for discussions. This work was supported in part by the NSFC under Grants
No. 11690034 and No.12075084; the Project supported by the Research Foundation of Education Bureau of Hunan Province, China under Grant No.20B371.

\end{acknowledgments}

\end{document}